\documentclass{article}
\usepackage[utf8]{inputenc}

\usepackage{array,multirow}
\usepackage{graphicx}
\usepackage{subfig}
\usepackage{amssymb,amsmath}
\usepackage{bm}
\usepackage{siunitx}
\usepackage{etoolbox}
\usepackage[T1]{fontenc}

\usepackage{pgfplots}
\usetikzlibrary{fit}
\pgfplotsset{compat=1.11,
        /pgfplots/ybar legend/.style={
        /pgfplots/legend image code/.code={%
        \draw[##1,/tikz/.cd,bar width=10pt,yshift=-0.2em,bar shift=0pt]
                plot coordinates {(0cm,0.8em)};},
},
}

\usepackage{hyperref}
\usepackage{cleveref}

\usepackage{float}
\usepackage{cite}

\newcommand{\STAB}[1]{\begin{tabular}{@{}c@{}}#1\end{tabular}}

\usepackage{color, soul}
 
\usepackage{spconf}

\usepackage{caption}
\usepackage{subfig}

\title{Guided Variational Autoencoder for Speech Enhancement \\ With a Supervised Classifier}

\name{Guillaume Carbajal, Julius Richter, Timo Gerkmann \thanks{This work has been funded by the German Research Foundation (DFG) in the transregio project Crossmodal Learning (TRR 169) and ahoi.digital.}}
\address{Signal Processing (SP), Universität Hamburg, Germany \\
\{guillaume.carbajal, julius.richter, timo.gerkmann\}@uni-hamburg.de}

\begin{document}

\ninept

\maketitle

\begin{abstract}
Recently, variational autoencoders have been successfully
used to learn a probabilistic prior over speech signals, which
is then used to perform speech enhancement. However, variational autoencoders are trained on clean speech only, which results in a limited ability of extracting the speech signal from noisy speech compared to supervised approaches. In this paper, we propose to guide the variational autoencoder with a supervised classifier separately trained on noisy speech. The estimated label is a high-level categorical variable describing the speech signal (e.g. speech activity) allowing for a more informed latent distribution compared to the standard variational autoencoder.
We evaluate our method with different types of labels on real recordings of different noisy environments. Provided that the label better informs the latent distribution and that the classifier achieves good performance, the proposed approach outperforms the standard variational autoencoder and a conventional neural network-based supervised approach.

\end{abstract}


\section{Introduction}

The task of single-channel speech enhancement consists in recovering a speech signal from a mixture signal captured with one microphone in a noisy environment \cite{vincent_audio_2018}. Common speech enhancement approaches estimate the speech signal using a filter in the time-frequency domain to reduce the noise signal while avoiding speech artifacts \cite{hendriks_dft-domain_2013}. Under the Gaussian assumption, the optimal filter in the minimum mean square error sense requires estimating the signal variances \cite{breithaupt_cepstral_2007, fevotte_nonnegative_2009, gerkmann_unbiased_2012}.  

Supervised deep neural networks (DNNs) have demonstrated excellent performance in estimating the speech signal \cite{narayanan_ideal_2013, huang_joint_2015, williamson_complex_2016, luo_conv-tasnet:_2019}. However, supervised approaches require labeled data which originates from pairs of noisy and clean speech. These pairs can be created synthetically. However, since supervised approaches may not generalize well to unseen situations, a large amount of pairs is needed to cover various acoustic conditions, e.g. different noise types, reverberation and different signal-to-noise ratios (SNRs).

Recently, deep generative models based on the variational autoencoder (VAE)  have gained attention for learning the probability distribution of complex data \cite{kingma_introduction_2019}. VAEs have been used to learn a prior distribution of clean speech, and have been combined with an untrained non-negative matrix factorization (NMF) noise model to estimate the signal variances using a Monte Carlo expectation maximization (MCEM) algorithm \cite{bando_statistical_2018, leglaive_variance_2018}. However, since the VAE speech model is trained in an unsupervised manner on clean speech only, its ability of extracting speech characteristics from noisy speech is limited in low SNRs. This results in limited speech enhancement performance compared to supervised approaches in already-seen noisy environments \cite{bando_statistical_2018}.


To overcome this limitation, the VAE can be conditioned on an auxiliary variable that allows for a more informed probabilistic latent distribution
\cite{kingma_semi-supervised_2014}.
Kameoka et al. used a VAE conditioned on the speaker identity to inform the speech prior for multichannel speech separation \cite{kameoka_supervised_2019}. However, their approach can only separate speakers which are included in the training set.
As a result, their approach aims at speaker-dependent speech separation and not at speaker-independent speech enhancement.

In this work, we propose to guide the VAE with a classiﬁer fully decoupled from the VAE. The classiﬁer is trained separately in a supervised manner with pairs of noisy and clean speech. The estimated label is a high-level categorical variable describing the speech signal (e.g. speech activity). 
We show that the choice of label is crucial for the performance of the proposed guided VAE. In addition, we show that a noise-robust classifier is also required to outperform the standard VAE and a conventional supervised DNN-based approach.



The rest of this paper is organized as follows.  In Section \ref{sec:background} we summarize the background related to the VAE for speech enhancement. Section \ref{sec:proposed} describes our proposed approach. The experimental setup is described in Section \ref{sec:experimental} which is followed by the evaluation in Section \ref{sec:results}.


\section{Background} \label{sec:background}

\subsection{Mixture model and filtering}

In the time-frequency domain using the short time Fourier transform (STFT), the mixture signal $x_{nf} \in \mathbb{C}$ is the sum of the clean speech $s_{nf} \in \mathbb{C}$ and the noise $b_{nf} \in \mathbb{C}$:
\begin{align}
x_{nf} = \sqrt{g_n}\, s_{nf} + b_{nf},
\label{eq:mixture_model}
\end{align}
at time frame index $n \in [1, N]$ and frequency bin $f \in [1,F]$, where $N$ denotes the number of time frames and $F$ the number of frequency bins of the utterance. The scalar  $g_n \in \mathbb R_{+}$ represents a frequency-independent but time-varying gain providing some robustness with respect to the time-varying loudness of different speech signals \cite{leglaive_variance_2018}.

Under the Gaussian assumption, the clean speech $s_{nf}$ can be estimated in the minimum mean square error sense using the Wiener estimator:
\begin{equation}
\widehat{s}_{nf} = \frac{\widehat{g}_n \widehat{v}_{s, nf}}{\widehat{g}_n \widehat{v}_{s, nf} + \widehat{v}_{b, nf}} \, x_{nf},
\label{eq:wiener}
\end{equation}
where $\widehat{v}_{s, nf}$ and $\widehat{v}_{b,nf}$ are the estimated variances of the clean speech $s_{nf}$ and the noise $b_{nf}$, respectively. Under a local stationary assumption, short-time power spectra $|s_{nf}|^2$ and $|b_{nf}|^2$ are unbiased estimates of the signal variances \cite{liutkus_gaussian_2011}.

\subsection{Model M1: standard VAE as a \mbox{speech prior}}

\begin{figure}[b]
\centering
\begin{minipage}[b]{0.48\linewidth}
  \centering
  \centerline{\includegraphics[scale=1]{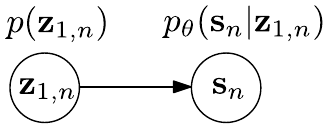}}
  \centerline{(a) Generative model}\medskip
\end{minipage}
\hfill
\begin{minipage}[b]{0.48\linewidth}
  \centering
  \centerline{\includegraphics[scale=1]{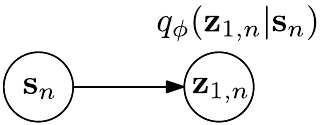}}
  \centerline{(b) Recognition model}\medskip
\end{minipage}
\caption{Model M1 consisting of (a) a generative model $p_\theta (\mathbf{s}_n|\mathbf{z}_{1,n}) p(\mathbf z_{1,n})$  and (b) a recognition model $q_\phi(\mathbf{z}_{1,n}|\mathbf{s}_n)$. \label{fig:model_M1}}
\end{figure}

The standard VAE which we refer to as model M1 is used to learn a prior over clean speech \cite{bando_statistical_2018, leglaive_variance_2018}. At time frame $n$, the frequency bins of clean speech $\mathbf s_{n} \in \mathbb C^F$ are modeled as
\begin{equation}
    \mathbf{s}_n | \mathbf{z}_{1,n} \sim \mathcal{CN}(\mathbf{0}, \text{diag}(\mathbf{v}_{\theta}(\mathbf{z}_{1,n}))), 
    \quad \mathbf{z}_{1,n} \sim \mathcal{N}(\mathbf 0, \mathbf I),
\end{equation}
where $\mathbf z_{1,n} \in \mathbb R^D$ denotes a latent variable of dimension $D$ and $\mathbf{v}_{\theta}: \mathbb{R}^D \mapsto \mathbb{R}_+^F$  represents a trainable feedforward DNN called the generative model or \emph{decoder}  pa\-ram\-e\-trized by $\theta$. (see Fig \ref{fig:model_M1}a).

In variational inference, the posterior of $\mathbf{z}_{1,n}$ is approximated as
\begin{equation}
    \mathbf{z}_{1,n} |\ \mathbf{s}_n \sim \mathcal N(\boldsymbol \mu_\phi(|\mathbf{s}_n|^2), \operatorname{diag}(\mathbf v_\phi(|\mathbf{s}_n|^2))) \label{eq:encoder_m1},
\end{equation}
where $\boldsymbol \mu_\phi: \mathbb R_+^F \mapsto \mathbb R^D$ and $\mathbf v_\phi: \mathbb R_+^F \mapsto \mathbb R_+^D$ represent feedforward  DNNs sharing the same input and hidden layers called the recognition model or \emph{encoder} which are parametrized by $\phi$ (see Fig \ref{fig:model_M1}b). Note that the absolute value and squaring in \eqref{eq:encoder_m1} are performed element-wise.

The generative model and recognition model are simultaneously trained by maximizing the evidence lower bound (ELBO) on the per-frame log-likelihood
\begin{align}
\begin{split}
      \log p_\theta(\mathbf{s}_n) \geq& \; \mathbb E_{q_\phi(\mathbf{z}_{1,n}| \mathbf{s}_n)} [\log p_\theta (\mathbf{s}_n| \mathbf{z}_{1,n})] \\
      & -  \mathcal{D}_\text{KL}(q_\phi (\mathbf{z}_{1,n}| \mathbf{s}_n)|| p(\mathbf{z}_{1,n})),
\end{split}
\end{align}
where the first term is the reconstruction loss and $ \mathcal{D}_\text{KL}(\cdot||\cdot)$ denotes the Kullback-Leibler divergence.

\subsection{Non-negative matrix factorization as noise model}

The noise variance is modeled with an untrained NMF as 
\begin{equation}
    v_{b,nf} = \left\{\mathbf H \mathbf W \right\}_{nf},
\end{equation}
where $\mathbf H \in \mathbb R_+^{N \times K}$ and $\mathbf W \in \mathbb R_+^{K \times F}$ are two non-negative matrices representing the temporal activations and spectral patterns of the noise power spectrogram. $K$ denotes the NMF rank.

\subsection{Clean speech estimation}

Given the speech prior provided by model M1 and the noise model, the mixture signal $x_{nf}$ is distributed as
\begin{equation}
x_{nf} |\ \mathbf{z}_{1,n} \sim \mathcal{CN}(0, {g}_n \{\mathbf v_{\theta} (\mathbf{z}_{1,n})\}_f + \{\mathbf H \mathbf W\}_{nf})
\label{eq:likelihood_mixture}
\end{equation}
where $\Theta_u = \{{g}_n, {\mathbf H}, {\mathbf W} \}$ are the unsupervised parameters to be estimated. Since the resulting optimization problem is intractable due to the non-linear relation between the speech variance and the latent variable, an MCEM algorithm is employed to iteratively optimize the unsupervised parameters $\Theta_u$ \cite{leglaive_variance_2018}. At each iteration, the estimated terms $\widehat{g}_n \{\mathbf v_{\theta} (\mathbf{z}_{1,n})\}_f$ and $\{\widehat{\mathbf H} \widehat{\mathbf W}\}_{nf}$ are supposed to get closer to the true variances $v_{s,nf}$ and $v_{b,nf}$, respectively, reaching a local optimum. Note that while the VAE operates on a frame-by-frame basis, the MCEM algorithm is offline, resulting in an offline estimation of the parameters.

 At test time, the recognition model takes the mixture signal $\mathbf{x}_{nf}$ as input instead of clean speech $\mathbf{s}_{nf}$. However, since model M1 is trained on clean speech only, its ability of extracting speech characteristics from the mixture $\mathbf{x}_{nf}$ is limited. As a result, the speech enhancement performance of the MCEM using M1 may be lower compared to supervised approaches trained on already-seen noisy environments \cite{bando_statistical_2018}.

\section{Guided variational autoencoder} \label{sec:proposed}

In this section, we propose a guided VAE which consists in an extension of model M1 combined with a supervised classifier. 

\subsection{Model M2: labeled VAE as a speech prior}

Inspired by Kingma et al.'s deep generative model for semi-supervised learning \cite{kingma_semi-supervised_2014}, we extend model M1 with a categorical variable $y_n \in \mathcal{Y}$ that characterizes a high-level feature of the speech signal (e.g. speech activity). Hereafter, we denote $y_n$ as the \emph{label}. The label is supposed to allow for a more informed probabilistic speech prior learned by the VAE.
We describe our choice for $y_n$ in Section \ref{sec:classifier}.

\begin{figure}[ht]
\centering
\begin{minipage}[b]{0.52\linewidth}
  \centering
  \centerline{\includegraphics[scale=1]{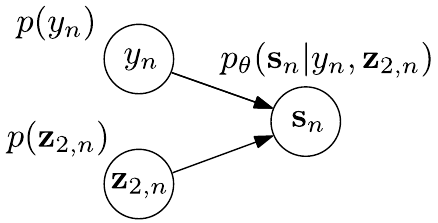}}
  \centerline{(a) Generative model}\medskip
\end{minipage}
\hfill
\begin{minipage}[b]{0.44\linewidth}
  \centering
  \centerline{\includegraphics[scale=1]{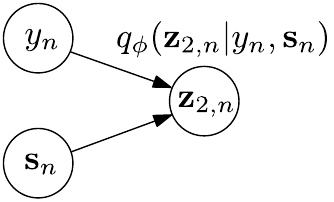}}
  \centerline{(b) Recognition model}\medskip
\end{minipage}

\caption{Model M2 consisting of (a) the guided generative model $p_\theta (\mathbf{s}_n| y_n, \mathbf{z}_{2,n}) p(\mathbf{z}_{2,n})p(y_n)$ and (b) the guided recognition model $q_\phi(\mathbf{z}_{2,n}| \mathbf{s}_n, y_n)$.}
\label{fig:model_M2}
\end{figure}

At time frame $n$, the frequency bins of clean speech $\mathbf s_{n}$ are generated as
\begin{equation}
    \mathbf{s}_n |  y_n, \mathbf{z}_{2,n} \sim \mathcal{CN}(\mathbf{0}, \text{diag}(\mathbf{v}_{\theta}(y_n, \mathbf{z}_{2,n}))),
\end{equation}
where $\mathbf z_{2,n}$ has the same prior as $\mathbf z_{1,n}$ and $\mathbf{v}_{\theta}: \mathcal{Y} \circ   \mathbb{R}^D \mapsto \mathbb{R}_+^F$ is a feedforward DNN resulting in the guided generative model (see Fig \ref{fig:model_M2}a). 
The  posterior of $\mathbf{z}_{2,n}$ is approximated as
\begin{equation}
    \mathbf{z}_{2,n} | y_n, \mathbf{s}_n \sim \mathcal N(\boldsymbol \mu_\phi(y_n,|\mathbf{s}_n|^2), \operatorname{diag}(\mathbf v_\phi(y_n,|\mathbf{s}_n|^2))) \label{eq:encoder_m2},
\end{equation}
where $\boldsymbol \mu_\phi: \mathcal{Y}  \circ \mathbb R_+^F  \mapsto \mathbb R^D$ and $\mathbf v_\phi: \mathcal{Y} \circ \mathbb R_+^F   \mapsto \mathbb R_+^D$ are feedforward DNNs sharing the same input and hidden layers resulting in the guided recognition model (see Fig \ref{fig:model_M2}b).


The guided generative model and recognition model are simultaneously trained by maximizing the ELBO on the per-frame joint log-likelihood 
\begin{align}
      &\log p_\theta(y_n, \mathbf{s}_n) \geq \; \mathbb E_{q_\phi(\mathbf{z}_{2,n}|y_n, \mathbf{s}_n)} [\log p_\theta (\mathbf{s}_n| y_n, \mathbf{z}_{2,n})] \notag \\
      & \quad\;\;\; -  \mathcal{D}_\text{KL}(q_\phi (\mathbf{z}_{2,n} | y_n, \mathbf{s}_n) || p(\mathbf{z}_{2,n})) + \log p (y_n),
\end{align}
where the first term is the reconstruction loss and $p(y_n)$ is the prior distribution of $y_n$. 

\subsection{Clean speech estimation}

\begin{figure}[t]
\centering
\includegraphics{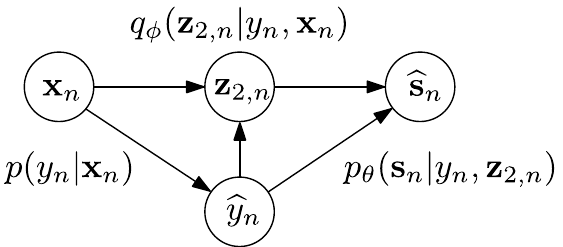}
\caption{Combined model M2 and classifier $p( y_n |\mathbf x_n)$ at test time.}
\label{fig:test_time}
\end{figure}

The estimation path at test time is shown in Fig. \ref{fig:test_time}. First, we use a classifier to  estimate $\widehat{y}_{n}$ from the mixture $\mathbf{x}_{n}$. Then, we use the mixture signal $\mathbf{x}_{n}$ and the estimated label $\widehat{y}_{n}$ as inputs for the guided recognition model. Given $\widehat{y}_{n}$ and the latent variable $\mathbf{z}_{2,n}$, the mixture signal $x_{nf}$ is distributed as
\begin{equation}
x_{nf} |  \widehat{y}_{n},  \mathbf{z}_{2,n} \sim \mathcal{CN}(0, {g}_n \{\mathbf v_{\theta} (\widehat{y}_{n}, \mathbf{z}_{2,n})\}_f 
+ \{\mathbf H \mathbf W\}_{nf})
\end{equation}

For estimating the unsupervised parameters we use the same MCEM configuration as for model M1. Provided that 1) the classifier is noise-robust and that 2) the label $y_{n}$ better informs the speech prior,
model M2 is supposed to better extract speech characteristics from the mixture $\mathbf{x}_{n}$ than model M1.

\subsection{Classifier for label estimation} \label{sec:classifier}

Since the classifier is fully decoupled from model M2, it can be trained separately. This fact can be used to construct the best classifier possible. In order to obtain a noise-robust classifier, we train a feedforward DNN in a supervised manner using the mixture power spectra $|\mathbf{x}_{n}|^2$ as inputs and corresponding labels $y_{n}$ as targets. The classifier outputs the posterior probability  $p(y_{n} | \mathbf{x}_{n})$ and the estimated label $\widehat{y}_{nf}$ is subsequently determined by selecting the class $c$ corresponding to the highest posterior probability $p(y_{n} = c | \mathbf{x}_{n})$.

We consider two types of labels related to speech activity. First, we use a classifier to perform voice activity detection (VAD), i.e. $y_n^\text{VAD}\in \{0, 1\}$. We use the binary cross entropy (BCE) as the learning objective for this classifier. The prior of $y_n^\text{VAD}$ is a symmetric Bernoulli distribution. Second, we consider a classifier to perform ideal binary mask (IBM) estimation, i.e. $\mathbf{y}_n^\text{IBM} \in \{0, 1\}^F$, which is equivalent to perform VAD per time-frequency bin. Thus, we use the BCE averaged over all frequency bins $f$ and the prior for each frequency bin is a symmetric Bernoulli distribution.

\section{Experimental setup} \label{sec:experimental}

\subsection{Dataset}


For training, we use the ``si\_tr\_s'' subset of the Wall Street Journal (WSJ0) dataset which consists of approximately $25\,\text{h}$ of clean speech \cite{garofolo_csr-i_1993}, and the noise signals DWASHING, NRIVER, OOFICE and TMETRO of the DEMAND dataset \cite{thiemann_demand_2013}. For validation, we use the ``si\_dt\_05'' subset of WSJ0 and the noise signals NFIELD, OHALLWAY, PSTATION and TBUS of the DEMAND dataset. All signals have a sampling rate of $16\, \text{kHz}$. For the test, we use the ``si\_et\_05'' subset of WSJ0 consisting of $651$ utterances, resulting in $1.5\,\text{h}$ and the noise signals from the "verification" subset of the QUT-NOISE dataset \cite{dean_qut-noise-sre_2015}, which we downsample to $16$ kHz. Note that both speakers and noise types in the test set are different than in the training set. Each mixture signal is created by uniformly sampling a noise type and mixing speech and noise signals at SNRs of $-5$, $0$ and $+5$ dB.

\subsection{Baselines}
Hereafter, we denote model M2 with VAD labels as \emph{M2+VAD} and model M2 with IBM labels as \emph{M2+IBM}. We use the DNN-based classifier described in Section \ref{sec:classifier} for the estimation of the VAD and IBM labels. To compare with our DNN-based IBM classifier, we also use a non-learned classifier which consists of the IBM estimator used inside the algorithm of Gerkmann and Hendriks \cite{gerkmann_unbiased_2012}, originally employed for noise PSD estimation.

For the baselines, we use model M1 and a feedforward DNN estimating a Wiener-like mask trained with the magnitude spectrum approximation loss \cite{weninger_discriminatively_2014}, which we denote as \emph{Supervised}.

\subsection{Hyperparameter settings}


 The STFT is computed using a $64\, \text{ms}$ Hann window with 75\% overlap, resulting in a frame period of $16\,\text{ms}$ and $F=513$ unique frequency bins. To obtain the ground truth for the VAD and IBM labels, we use the method of Heymann et al. related to clean speech \cite{heymann_neural_2016}.

For a fair comparison between all the approaches, we consider a similar architecture for each model. Tab. \ref{tab:model_configurations} shows the configuration of the models. In particular, we consider $5$ hidden layers for \emph{Supervised} to match the same number of layers as models M1 and M2 (\textit{encoder} + \textit{decoder}). Model M1 has \num[group-separator={,}]{171297} learnable parameters whereas \emph{M2+VAD} has \num[group-separator={,}]{177729} and \emph{M2+IBM} has \num[group-separator={,}]{302625}. The VAD classifier has \num[group-separator={,}]{82433} learnable parameters whereas the IBM classifier has \num[group-separator={,}]{148993} and \emph{Supervised} has \num[group-separator={,}]{198017}.

We use the Adam optimizer with standard configuration and a learning rate of $10^{-3}$ \cite{kingma_adam_2014}.  We set the batch size to $128$. Note that because the learning objective of the classifier is scale-dependent, the DNN input $|\mathbf{x}_{n}|^2$ needs to be normalized at training time. This is not the case for models M1 and M2 since the reconstruction loss (i.e. the Ikatura-Saito distance) is scale-independent. Early stopping with a patience of $20$ epochs is performed using the validation set. For the MCEM we follow the settings of Leglaive et al. and set the NMF rank to $K=10$ \cite{leglaive_variance_2018}. For the non-learned IBM classifier, we use the standard configuration of the IBM estimator as in Gerkmann and Hendriks \cite{gerkmann_unbiased_2012}.



\begin{table}[h]
\centering
\sisetup{table-align-uncertainty=true,separate-uncertainty=true,}
\renewrobustcmd{\bfseries}{\fontseries{b}\selectfont}
\renewrobustcmd{\boldmath}{}
\begin{tabular}{l | c c c | c}
& \multicolumn{3}{c|}{Hidden layers} & Output layer \\
\textbf{Model} & \textbf{\# layers} & \textbf{\# units} & \textbf{act. fn} & \textbf{act. fn} \\
\hline
\textit{Encoder} & $2$ & $128$ & $\operatorname{tanh}$ & $\operatorname{identity}$ \\
\textit{Decoder} & $2$ & $128$ & $\operatorname{tanh}$ & $\operatorname{exp}$ \\
DNN classifer & $2$ & $128$ & $\operatorname{ReLU}$ & $\operatorname{sigmoid}$  \\
\emph{Supervised} & $5$ & $128$ & $\operatorname{ReLU}$ & $\operatorname{sigmoid}$
\end{tabular} 
\caption{Model configurations.}
\label{tab:model_configurations}
\end{table}

\subsection{Metrics}

\begin{table*}[t]
\centering
\sisetup{table-align-uncertainty=true,separate-uncertainty=true,}
\renewrobustcmd{\bfseries}{\fontseries{b}\selectfont}
\renewrobustcmd{\boldmath}{}
\vspace{-1em}
\begin{tabular}{l c | c c | c c c}
\multicolumn{2}{c}{} &  \multicolumn{2}{c}{} & \multicolumn{3}{c}{\textbf{Input SNR}} \\
\textbf{Model} & \textbf{Classifier} & \textbf{F1-score} & \textbf{SI-SDR} &
$-5\,\text{dB}$& $0\,\text{dB}$  &  $5\,\text{dB}$ \\
 \hline
Mixture & -- & \multicolumn{1}{c}{--} & $\;\;\;0.2 \pm 0.3$ &  $-5.0 \pm 0.0$ &  $\;\;\;0.0 \pm 0.0$ & $\;\;5.0 \pm 0.0$\\  
 \hline
\emph{Supervised} & -- & \multicolumn{1}{c}{--} & $\;\;\;4.7 \pm 0.4$ & $-0.9 \pm 0.5$ &$\;\;\;4.5 \pm 0.4$ & $\;\;9.7 \pm 0.3$\\  
\hline
\emph{M1} & -- & \multicolumn{1}{c}{--} & $\;\;\;6.4 \pm 0.4$  &  $\;\;\;1.4 \pm 0.6$ & $\;\;\;6.3 \pm 0.5$ & $10.8 \pm 0.4$\\  
\hline
\multirow{2}{*}{\emph{M2+VAD}} &  DNN & $0.82$ & $\;\;\;6.3 \pm 0.6$ &$\;\;\;1.3 \pm 0.6$ & $\;\;\;6.3 \pm 0.5$ & $10.7 \pm 0.4$ \\ 
\cline{2-7}
 & oracle & $1.00 $ & $\;\;\;6.8 \pm 0.4$ &$\;\;\;2.0 \pm 0.6$ & $\;\;\;6.8 \pm 0.5$ & $11.2 \pm 0.4$ \\
\hline
\multirow{3}{*}{\emph{M2+IBM}} & \cite{gerkmann_unbiased_2012} & $0.34$ & $-0.5 \pm 0.3$ & $-2.2 \pm 0.6$ & $-0.2 \pm 0.5$ & $\;\;0.8 \pm 0.6$\\
 & DNN & $0.62$ & $\;\;\;\textbf{7.3} \pm \textbf{0.4}$ & $\;\;\;\textbf{2.8} \pm \textbf{0.6}$ & $\;\;\;\textbf{7.1} \pm \textbf{0.5}$ & $\textbf{11.5} \pm \textbf{0.4}$\\  
\cline{2-7}
 & oracle & $1.00 $ & $\;\;\;9.5 \pm 0.4$ & $\;\;\;5.7 \pm 0.5$ & $\;\;\;9.3 \pm 0.5$ & $13.3 \pm 0.4$ \\ 
\end{tabular}
\caption{Average SI-SDR (in dB) and  95\% confidence intervals on the test set on average and for different input SNRs.}
\label{tab:results}
\end{table*} 

To evaluate the classification performance of the different classifiers, we use the F1-score which combines the precision and recall rates. To evaluate the speech enhancement performance of the approaches, we use the scale-invariant signal-to-distortion ratio (SI-SDR) measured in dB \cite{roux_sdr_2019}.

\section{Results} \label{sec:results}

Tab. \ref{tab:results} shows the results on average and per input SNR. Regarding the results on average, \emph{M2+VAD} with DNN classifier outperforms \emph{Supervised} by $1.7\,\text{dB}$ but is outperformed by \emph{M1} by $0.1\,\text{dB}$. \emph{M2+VAD} with the oracle classifier outperforms \emph{M1} by $0.4\,\text{dB}$ but the difference is not statistically significant. We conclude that, even with the best classifier, VAD does not inform the speech prior learned by the VAE statistically better.

\emph{M2+IBM} with DNN classifier outperforms \emph{Supervised} by $2.6\,\text{dB}$ and \emph{M1} by $0.9\,\text{dB}$ on average. The performance of \emph{M2+IBM} is also statistically significant compared to these two models. However, the performance of \emph{M2+IBM} with the non-learned classifier dramatically drops compared to the DNN classifier. Since the classification performance of the non-learned classifier is worse than the DNN classifier, we conclude that the performance of \emph{M2+IBM} crucially depends on the classifier. Finally, the performance of \emph{M2+IBM} with the oracle classifier shows that IBM informs the speech prior significantly better.
Thus, the performance of \emph{M2+IBM} could be improved with a better classifier.

\emph{M2+IBM} with DNN classifier also outperforms \emph{M1} for all input SNRs. The difference between the two models gets larger as the input SNR decreases. Therefore, \emph{M2+IBM} with the DNN classifier is particularly more robust to noise than \emph{M1} in low SNRs.


From informal listening tests, we can state that \emph{M2+IBM} with DNN classifier typically reduces the noise better than \emph{M1}, which is particularly obvious in the presence of nonstationary noise and transient interferences such as bursts. Code and audio examples are available online\footnote{\url{https://uhh.de/inf-sp-guided2021}}.

\section{Conclusion} \label{sec:conclusion}

We proposed to guide a VAE for speech enhancement with a supervised classifier separately trained on noisy speech. We evaluated our method with labels corresponding to VAD and IBM on real recordings of different noisy environments. Using the IBM as label and a feedforward DNN classifier, the guided VAE outperforms the standard VAE and a feedforward DNN-based Wiener filter, particularly in low SNRs. Improving the classifier by taking time dependencies and/or visual information into account could further improve the guided VAE. The model could then be compared to other deep generative models taking temporal dependencies into account \cite{richter_speech_2020}.
\bibliographystyle{ieeetr}
\bibliography{zotero-refs}

\begin{thebibliography}{10}

\bibitem{vincent_audio_2018}
E.~Vincent, T.~Virtanen, and S.~Gannot, eds., {\em Audio {{Source Separation}}
  and {{Speech Enhancement}}}.
\newblock {Hoboken, NJ}: {John Wiley \& Sons}, 2018.

\bibitem{hendriks_dft-domain_2013}
R.~C. Hendriks, T.~Gerkmann, and J.~Jensen, {\em {{DFT}}-{{Domain Based
  Single}}-{{Microphone Noise Reduction}} for {{Speech Enhancement}}: A
  {{Survey}} of the {{State}}-of-the-{{Art}}}.
\newblock No.~11 in Synthesis Lectures on Speech and Audio Processing,
  {Williston, VT}: {Morgan \& Claypool}, 2013.

\bibitem{breithaupt_cepstral_2007}
C.~Breithaupt, T.~Gerkmann, and R.~Martin, ``Cepstral {{Smoothing}} of
  {{Spectral Filter Gains}} for {{Speech Enhancement Without Musical Noise}},''
  {\em IEEE Signal Processing Letters}, vol.~14, pp.~1036--1039, Dec. 2007.

\bibitem{fevotte_nonnegative_2009}
C.~F{\'e}votte, N.~Bertin, and J.-L. Durrieu, ``Nonnegative {{Matrix
  Factorization}} with the {{Itakura}}-{{Saito Divergence}}: {{With
  Application}} to {{Music Analysis}},'' {\em Neural Computation}, vol.~21,
  pp.~793--830, Mar. 2009.

\bibitem{gerkmann_unbiased_2012}
T.~Gerkmann and R.~C. Hendriks, ``Unbiased {{MMSE}}-{{Based Noise Power
  Estimation With Low Complexity}} and {{Low Tracking Delay}},'' {\em IEEE
  Transactions on Audio, Speech, and Language Processing}, vol.~20,
  pp.~1383--1393, May 2012.

\bibitem{narayanan_ideal_2013}
A.~Narayanan and D.~Wang, ``Ideal ratio mask estimation using deep neural
  networks for robust speech recognition,'' in {\em {{ICASSP}}},
  pp.~7092--7096, May 2013.

\bibitem{huang_joint_2015}
P.~Huang, M.~Kim, M.~{Hasegawa-Johnson}, and P.~Smaragdis, ``Joint
  {{Optimization}} of {{Masks}} and {{Deep Recurrent Neural Networks}} for
  {{Monaural Source Separation}},'' {\em IEEE/ACM Transactions on Audio,
  Speech, and Language Processing}, vol.~23, pp.~2136--2147, Dec. 2015.

\bibitem{williamson_complex_2016}
D.~S. Williamson, Y.~Wang, and D.~Wang, ``Complex {{Ratio Masking}} for
  {{Monaural Speech Separation}},'' {\em IEEE/ACM Transactions on Audio,
  Speech, and Language Processing}, vol.~24, pp.~483--492, Mar. 2016.

\bibitem{luo_conv-tasnet:_2019}
Y.~Luo and N.~Mesgarani, ``Conv-{{TasNet}}: {{Surpassing Ideal
  Time}}\textendash{{Frequency Magnitude Masking}} for {{Speech Separation}},''
  {\em IEEE/ACM Transactions on Audio, Speech, and Language Processing},
  vol.~27, pp.~1256--1266, Aug. 2019.

\bibitem{kingma_introduction_2019}
D.~P. Kingma and M.~Welling, ``An {{Introduction}} to {{Variational
  Autoencoders}},'' {\em Foundations and Trends in Machine Learning}, vol.~12,
  no.~4, pp.~307--392, 2019.

\bibitem{bando_statistical_2018}
Y.~Bando, M.~Mimura, K.~Itoyama, K.~Yoshii, and T.~Kawahara, ``Statistical
  {{Speech Enhancement Based}} on {{Probabilistic Integration}} of
  {{Variational Autoencoder}} and {{Non}}-{{Negative Matrix Factorization}},''
  in {\em {{ICASSP}}}, pp.~716--720, Apr. 2018.

\bibitem{leglaive_variance_2018}
S.~Leglaive, L.~Girin, and R.~Horaud, ``A variance modeling framework based on
  variational autoencoders for speech enhancement,'' in {\em {{MLSP}}},
  pp.~1--6, Sept. 2018.

\bibitem{kingma_semi-supervised_2014}
D.~P. Kingma, S.~Mohamed, D.~Jimenez~Rezende, and M.~Welling, ``Semi-supervised
  learning with deep generative models,'' in {\em {{NeurIPS}}}, pp.~3581--3589,
  {Curran Associates, Inc.}, 2014.

\bibitem{kameoka_supervised_2019}
H.~Kameoka, L.~Li, S.~Inoue, and S.~Makino, ``Supervised determined source
  separation with multichannel variational autoencoder,'' {\em Neural
  Computation}, vol.~31, pp.~1891--1914, Sept. 2019.

\bibitem{liutkus_gaussian_2011}
A.~Liutkus, R.~Badeau, and G.~Richard, ``Gaussian {{Processes}} for
  {{Underdetermined Source Separation}},'' {\em IEEE Transactions on Signal
  Processing}, vol.~59, pp.~3155--3167, July 2011.

\bibitem{garofolo_csr-i_1993}
J.~S. Garofolo, D.~Graff, D.~Paul, and D.~S. Pallett, {\em {CSR-I (WSJ0)
  Sennheiser.}}
\newblock 1993.

\bibitem{thiemann_demand_2013}
J.~Thiemann, N.~Ito, and E.~Vincent, ``{{DEMAND}}: {{A Collection Of
  Multi}}-{{Channel Recordings Of Acoustic Noise In Diverse Environments}},''
  June 2013.

\bibitem{dean_qut-noise-sre_2015}
D.~Dean, A.~Kanagasundaram, H.~Ghaemmaghami, H.~Rahman, and S.~Sridharan, ``The
  {{QUT}}-{{NOISE}}-{{SRE Protocol}} for the {{Evaluation}} of {{Noisy Speaker
  Recognition}},'' in {\em Interspeech}, pp.~3456--3460, 2015.

\bibitem{weninger_discriminatively_2014}
F.~Weninger, J.~R. Hershey, J.~L. Roux, and B.~Schuller, ``Discriminatively
  trained recurrent neural networks for single-channel speech separation,'' in
  {\em {{GlobalSIP}}}, pp.~577--581, Dec. 2014.

\bibitem{heymann_neural_2016}
J.~Heymann, L.~Drude, and R.~{Haeb-Umbach}, ``Neural network based spectral
  mask estimation for acoustic beamforming,'' in {\em {{ICASSP}}},
  pp.~196--200, Mar. 2016.

\bibitem{kingma_adam_2014}
D.~P. Kingma and J.~Ba, ``Adam: {{A Method}} for {{Stochastic Optimization}},''
  in {\em {{ICLR}}}, Dec. 2014.

\bibitem{roux_sdr_2019}
J.~L. Roux, S.~Wisdom, H.~Erdogan, and J.~R. Hershey, ``{{SDR}} \textendash{}
  {{Half}}-baked or {{Well Done}}?,'' in {\em {{ICASSP}}}, pp.~626--630, May
  2019.

\bibitem{richter_speech_2020}
J.~Richter, G.~Carbajal, and T.~Gerkmann, ``Speech {{Enhancement}} with
  {{Stochastic Temporal Convolutional Networks}},'' in {\em Interspeech}, Oct.
  2020.

\end{thebibliography}
\end{document}